# On Accuracy of Community Structure Discovery Algorithms

[1,2]Günce Keziban Orman, [1]Vincent Labatut, [2]Hocine Cherifi
[1,]Galatasaray University, korman@gsu.edu.tr, vlabatut@gsu.edu.tr.
[2,] University of Burgundy, hocine.cherifi@u-bourgogne.fr

*Abstract*

*Community structure discovery in complex networks is a quite challenging problem spanning many applications in various disciplines such as biology, social network and physics. Emerging from various approaches numerous algorithms have been proposed to tackle this problem. Nevertheless little attention has been devoted to compare their efficiency on realistic simulated data. To better understand their relative performances, we evaluate systematically eleven algorithms covering the main approaches. The Normalized Mutual Information (NMI) measure is used to assess the quality of the discovered community structure from controlled artificial networks with realistic topological properties. Results show that along with the network size, the average proportion of intra-community to inter-community links is the most influential parameter on performances. Overall, "Infomap" is the leading algorithm, followed by "Walktrap", "SpinGlass" and "Louvain" which also achieve good consistency.*

## 1. Introduction

Complex networks have become a very popular modeling tool during the last decade. They allow studying a given system by representing its components and their relationships with nodes and links, respectively [1]. Complex network analysis helps surfacing the studied system properties which could not be discovered at first sight. One of the most prominent sub-domains in complex network analysis is community detection. A community is a cohesive subset of nodes with denser inner links, relatively to the rest of the network [2]. A community structure is a set of communities, or more precisely a partition of the network node set. This popular research topic has applications in many fields such as biology, social science, physics, computer science, business science, etc. Tens of algorithms have been proposed to deal with community structure discovery. They are based on a whole range of principles such as hierarchical clustering, optimization methods, graph partitioning, spectral properties of the network, and others.

Authors traditionally test their community detection algorithms on benchmark graphs [3,4], artificial or from the real-world. However, limiting these tests to real-world networks can be considered as an issue for several reasons. First, obtaining such networks is a costly and difficult task, and determining reference communities can be done only by experts. The benchmarks consequently contain only a few graphs, and their communities are not always defined objectively. Second, a complex network is characterized by various topological properties such as its average degree, degree distribution, shortest average path, etc. As it is not possible to control these features in a real-world network the algorithm is tested on a very specific and limited set of features. Artificial networks seem to overcome these limitations, because it is possible to randomly generate many of them, while controlling their properties. All that is needed is a generative model able to produce networks with features similar to those of real-world networks. Of course, artificial networks must not be seen as a substitute to real-world networks, but rather as a complement. The most popular generative model used to test community detection algorithms, has been defined by Newman and Girvan [5]. However, although widely used for comparative purposes [2,5-7], it is limited in terms of realism [8]. Indeed it generates small networks with equal size communities and the degrees of the nodes are approximately the same. To overcome these drawbacks several variants producing larger networks and communities with heterogeneous sizes have been defined. Although these implement some improvements on certain properties, they still lack some important real-world properties like power-law distributed community sizes. More recently, a different approach appeared, based on some rewiring process [8,9]. It increased the realism level even more by producing networks with power-law distributed degree. Among these works, the model proposed by Lancichinetti *et al*. [8], exhibits the most realistic properties. Indeed it is able to generate networks with controlled power-law degree and power-law community size distributions. This new benchmark graph allows revisiting the problem of community structure discovery algorithms comparison. Indeed, until now, there is no agreement about a set of reliable algorithms that one can use in applications. Furthermore little is known about the relationship between performances and networks parameters. This work aims to provide some answers on these issues of primary importance for applications. To do so, we explore a wide range of community detection methods originating from different approaches on a set of artificial networks with various size and topological properties. We use the normalized mutual information measure [10] to quantify the similarity between the original generated community structure and the one estimated by the algorithms and to assess the quality of algorithms under investigation.

The rest of the paper is organized as follows. In section 2, we give an overview of the main families of community detection algorithms and present briefly some representative ones selected in order to perform our comparison. In section 3, we describe the network generation model used to create the artificial networks, and the normalized mutual information measure which allows us to assess the algorithms performances. In section 4, we present the results of the comparative evaluation and discuss the algorithms performances. We conclude in section 5 with some general observations.

## 2. Community Detection Algorithms

Identifying communities in a network is an important issue for many real-world applications in various scientific fields. Over the years, many methods have been devised to provide efficient community discovery algorithms. As the spectrum is wide, building taxonomy of solutions is not easy. They can be classified in different ways, and depending on the selected criteria, one algorithm can belong to more than one category. Here, we choose to focus on the process implemented by the algorithms. We consider this as their main characteristic, since it directly affects the nature of the detected communities. As a result, we group the algorithms in six different categories. In this section, we describe these categories and the representative set of algorithms we selected.

### 2.1. Link-Centrality-Based Algorithms

The algorithms based on link-centrality measures rely on a hierarchical divisive approach. Initially the whole network is seen as a single community, i.e. all nodes are in the same community. The most central links are then repeatedly removed. The underlying assumption is that these particular links are located between the communities. After a few steps, the network is split in several components which can be considered as communities in the initial network. Iterating the process, one can split each discovered community again, resulting in a finer community structure. This eventually leads to a network in which each node is isolated, and therefore constitutes its own community. By considering the communities detected at each step of the process, one obtains a hierarchy of community structures. The choice of the best one is generally performed using a measure estimating the quality of community structures, such as the modularity [2].

Algorithms of this category differ in the way they select the links to be removed. The first and most known algorithm using this approach was proposed by Newman [5], and relies on the *edge-betweenness* measure. It estimates the centrality of a link by considering the proportion of shortest paths going through it in the whole network. As the complexity of this algorithm is high, it is not well suited for very large networks.

Radicchi et *al*. proposed a variation called *Radetal* [11], based on *link transitivity* instead of edge-betweenness. This measure is defined as the number of triangles to which a given link belongs, divided by the number of triangles that might potentially include it. Its lower complexity makes it more appropriate for large networks. It is used as the representative of the link centrality based approach and will be referred as RA in the following.

## 2.2. Modularity Optimization Algorithms

Modularity is a prominent measure of the quality of a community structure introduced by Newman and Girvan [2]. It measures internal connectivity of identified communities with reference to a randomized null model with the same degree distribution. Modularity optimization algorithms try to find the best community structure in terms of modularity. They diverge on the optimization process they are based on. As this approach is very influential in the community detection literature we consider three algorithms for investigation.

*FastGreedy* (FG) developed by Newman *et al.* [6] relies on a greedy optimization method applied to a hierarchical agglomerative approach. The agglomerative approach is symmetrical to the divisive one described in the previous subsection. In the initial state, each node constitutes its own community. The algorithm merges those communities step by step until only one remains, containing all nodes. The greedy principle is applied at each step, by considering the largest increase (or smallest decrease) in modularity as the merging criterion. Because of its hierarchical nature, FG produces a hierarchy of community structures like the divisive approaches. The best one is selected by comparing their modularity values.

*Louvain* (LV) is another optimization algorithm proposed by Blondel et *al.* [12]. It is an improvement of FG, introducing a two-phase hierarchical agglomerative approach. During the first phase, the algorithm applies a greedy optimization to identify the communities. During the second phase, it builds a new network whose nodes are the communities found during the first phase. The intra-community links are represented by self-loops, whereas the inter-community links are aggregated and represented as links between the new nodes. The process is repeated on this new network, and stops when only one community remains.

*Spinglass* (SG) by Reichardt and Bornholdt [13] relies on an analogy between a very popular statistical mechanic model called *Potts spin glass,* and the community structure. It applies the simulated annealing optimization technique on this model to optimize the modularity.

## 2.3. Spectral Algorithms

Spectral algorithms take advantage of various matrix representations of networks. Classic spectral graph partitioning techniques focus on the eigenvectors of the Laplacian matrix. They were designed to find the partition minimizing the links lying in-between node groups. However, these methods were designed for slightly different contexts (e.g. user-specified number of communities). For real-world complex networks, the community number is unknown. Thus, these methods are not efficient in our case. The methods we selected are variants adapted to complex networks analysis.

*Leading Eigenvector* (LEV) is proposed by Newman [14]. It applies the classic graph partitioning approach, but to a so-called modularity matrix instead of the Laplacian. Doing so, it performs an optimization of the modularity instead of the objective measures used in classic graph partitioning, such as the minimal cut.

*Commfind* (CF) is developed by Donetti and Muñoz [15]. It combines the analysis of the Laplacian matrix eigenvectors used in classic graph partitioning with a cluster analysis step. Instead of using the best eigenvector to iteratively perform bisections of the network, it takes advantage of the $m$ best ones. Communities are obtained by a cluster analysis of the projected nodes in this $m$-dimensional space.

## 2.4. Random-Walk-Based Algorithms

Several algorithms use random walks in various ways to partition the network into communities. We retain two of them in our comparisons.

*Walktrap* (WT) by Pons and Latapy [7] uses a hierarchical agglomerative method like FG but with a different merging criterion. Unlike FG, which relies on the modularity measure, WT uses a node-to-node distance measure to identify the closest communities. This distance is based on the concept of random-walk. If two nodes are in the same community, the probability to get to a third one located in the same community through a random walk should not be very different for both of them. The distance is constructed by summing these differences over all nodes, with a correction for degree.

*MarkovCluster* (MCL) simulates a diffusion process in the network to detect communities [16]. This method relies on the network *transfer matrix*, which describes the transition probabilities for a random walker evolving in this network. Two transformations; expansion and inflation are iteratively applied on this matrix until convergence. Expansion raises the transfer matrix to a power $p$. The result is a matrix showing the probability for a random walker to start from node $i$ and reach node $j$ in $p$ steps. Inflation consists in raising each element in the matrix to some specified power, in order to favor the higher probability values. These correspond to nodes presumably belonging to the same community. The resulting matrix is then normalized to get a new transfer matrix, and the process is repeated until convergence. The final matrix can be interpreted as the adjacency matrix of a network with disconnected components, which correspond to communities in the original network.

## 2.5. Information-Based Algorithms

Information-Based algorithms use tools derived from the information theory to estimate the best partition of the network. The main idea of those approaches is to take advantage of the community structure in order to represent the network using less information than that encoded in the full adjacency matrix. We selected two algorithms from this category.

*Infomod* (IND) was proposed by Rosvall and Bergstorm [17]. It is based on a simplified representation of the network focusing on the community structure: a community matrix and a membership vector. The former is an adjacency matrix defined at the level of the communities (instead of the nodes), and the latter associates each node to a community. The authors use the mutual information measure to quantify the amount of information from the original network contained in the simplified representation. They obtain the best partition by considering the representation associated to the maximal mutual information.

*Infomap* (INP) is another algorithm developed by Rosvall and Bergstorm [18]. The community structure is represented through a two-level nomenclature based on Huffman coding: one to distinguish communities in the network and the other to distinguish nodes in a community. The problem of finding the best partition is expressed as minimizing the quantity of information needed to represent some random walk in the network using this nomenclature. With a partition containing few inter-community links, the walker will probably stay longer inside communities, therefore only the second level will be needed to describe its path, leading to a compact representation. The authors optimize their criterion using simulated annealing.

## 2.6. Other Algorithms

A number of algorithms do not fit in the previously described approaches. We selected the *Label Propagation* (LP) algorithm by Raghavan *et al.* [19], which uses the concept of node neighborhood and simulates the diffusion of some information in the network to identify communities. Initially, each node is labeled with a unique value. Then an iterative process takes place, where each node takes the label which is the most spread in its neighborhood (ties are broken randomly). This process goes on until convergence, i.e. each node has the majority label of its neighbors. Communities are then obtained by considering groups of nodes with the same label. By construction, one node has more neighbors in its community than in the others.

## 3. Method

### 3.1. Artificial Network Generation

In many community detection works, artificial networks with a community structure are generated using models comparable to Newman and Girvan's [5,6]. It produces networks with a degree following a Poisson distribution. Yet, it is well known that in most real-world networks, the degree follows a power-law distribution [1]. Networks with this property are called scale-free, because the shape of their degree distribution does not depend on their size (some other properties may, though). Moreover, in

Newman and Girvan's approach, all the communities have the same size, whereas in real-world networks, the community size is supposed to follow a power-law distribution too.

Lancichinetti *et al*. proposed a new class of benchmark graph to generate undirected and unweighted networks with mutually exclusive communities [8]. We use the abbreviation LFR to refer to this model in the following. In the produced networks, nodes degrees and community sizes are both power-law distributed. Moreover, this method allows controlling directly the following parameters: number of nodes $n$, desired average $\langle k \rangle$ and maximum $k_{max}$ degrees, exponent $\gamma$ for the degree distribution, exponent $\beta$ for the community size distribution, and mixing coefficient $\mu$. The latter represents the desired average proportion of links between a node and nodes located outside its community, called inter-community links. Consequently, the proportion of intra-community links is $1-\mu$. It is generally not possible to meet this constraint exactly, and the mixing coefficient value is therefore only approximated in practice. It is an important parameter, because it determines how clearly the communities are defined in terms of structure. For small values, the communities are distinctly separated, whereas for high values, the network has almost no community structure, making community identification a difficult task. There is a limit value ($\mu_{lim}$,) above which LFR cannot produce networks with a significant community structure (there is more inter-community than intra-community links). This limit depends on the number of nodes in the whole network ($n$) and in the largest community ($\max(n_{c_i})$) [20]:

$$\mu_{lim} = \frac{(n - \max(n_{c_i}))}{n} \tag{1}$$

The LFR algorithm proceeds in three-steps. First, it uses the configuration model to generate a network with average degree $\langle k \rangle$, maximum degree $k_{max}$ and power-law degree distribution with exponent $\gamma$. Second, the nodes are affected to the communities so that their sizes follow a power-law distribution with exponent $\beta$. Third, an iterative process takes place to randomly rewire certain links, so that $\mu$ is approximated, but without changing the degree distribution.

### 3.2. Performance Assessment

The assessment of the quality must be reliable in order to compare efficiently the communities detected by the tested algorithms. We use artificial networks, whose communities are known *a priori*. In this context, it is possible to quantify the match between the actual and estimated communities through several different measures. Normalized mutual information (NMI) is a recent measure used in the context of classic cluster analysis, to compare two different partitions of the same data set [10]. Because of its widespread use and efficiency, we selected this measure to compare algorithms in this study. The measure is derived from a confusion matrix whose element $m_{ij}$ represents the number of nodes classified in community $i$ by the considered algorithm, when they actually belong to community $j$. This matrix is usually rectangular, because the algorithm does not necessary estimate the correct number of communities.

$$NMI = \frac{-2 \sum_i \sum_j m_{ij} \log(n m_{ij} / (\sum_j m_i \times \sum_i m_j))}{\sum_i m_{i+} \log(\sum_j m_i / n) + \sum_j m_{+j} \log(\sum_i m_j / n)} \tag{2}$$

NMI ranges from 0 to 1. If the estimated communities correspond perfectly to the actual ones, the measure takes the value 1. It is zero when the estimated communities are independent from the actual ones.

### 4. Results & Discussion

We generate a collection of networks using the LFR model. The parameter values we use are typical of measurements on real-world networks. Experimental studies show the $\gamma$ coefficient usually ranges from 2 to 3 [1]. The power-law parameter $\beta$ for the community size distribution is known to range from 1 to 2 [2]. The average and maximal degrees generally depend on the number of nodes in the network. For a scale-free network, it is estimated to be $\langle k \rangle \sim k_{max}^{-\gamma+2}$ and $k_{max} \sim n^{1/(\gamma-1)}$ [1] respectively. The parameter values used in our experiments are indicated in Table 1.

Table 1. Network generation parameters

| Parameter | Value |
|---|---|
| Node Number ($n$) | {100, 500, 1000, 5000} |
| Average Degree ($\langle k \rangle$) | {5, 15, 30} |
| Maximum Degree ($k_{max}$) | $3 * \langle k \rangle$ |
| Degree Distribution exponent ($\gamma$) | {2, 3} |
| Community Size Distribution Exponent ($\beta$) | {1, 2} |
| Mixing Coefficient (μ) | [0.05; 0.95] with a 0.05 step |

For each combination of parameters, we produce a sample containing 25 networks. We test the eleven community detection algorithms presented in section 2 on all generated samples and assess their partitioning performance using NMI. To compare them, we average the measured NMI values over the 25 networks of each sample. We then analyze those results relatively to each parameter taken individually. The performance of the algorithms does not change considerably for different values of $\gamma$ and $\beta$. To assess their influence, we calculate Pearson's correlation between each parameter and the NMI values. According to these correlation values, the performance of the algorithms are not linked to $\beta$ and $\gamma$. Indeed the correlation values are always less than 0.06 in absolute value. This is not the case for the mixing coefficient which is highly correlated to the performances, (correlations above 0.5). The average and maximum degrees as well have non-negligible influence if we refer to the correlation values (above 0.3).

As expected, the parameter with the strongest effect on the algorithms performance is $\mu$. This parameter directly affects the community structure. One would expect a good algorithm to find a relatively correct community structure until $\mu_{lim}$. In Figure 1, we represent the NMI variation as a function of mixing coefficient value for networks of size 5000. The largest observed community size is 700 for these networks, thus $\mu_{lim} = 0.86$. One can distinguish two distinct parts in this plot, separated by $\mu_{lim}$. Above this value the generated networks have no community structure, and are therefore inappropriate for our study. Let's focus on the values lower than $\mu_{lim}$. For all the algorithms the performance decreases when the mixing coefficient increase. We can distinguish three different type of behavior according to the shape of variation (reverse sigmoid, linear, and irregular). Except for *Radetal, Fast Greedy* and *Commfind*, the curves representing the performance of the algorithms take the form of reverse sigmoid, with different slopes and inflection points. Among them, *Infomap* performs very well for a wide range of mixing coefficient value almost until it reaches $\mu_{lim}$. This algorithm gives the best results in terms of NMI. *Walktrap*, *Spinglass*, and *Label Propagation* also display good, although lower performances. *Louvain* performance starts to decrease before the other algorithms, but it is still efficient. *MarkovCluster* performances start to decrease when $\mu$ gets close to 0.4. But, despite this behavior, the NMI value stays relatively high (around 0.8), even when $\mu$ gets closer to 1. We observe more detailed the result of *MarkovCluster* for the second zone when the community structure is not well defined. We see that this algorithm finds many small communities including only one or two nodes. Although it does not finds appropriate communities with reference structure, the NMI values are quite high. This can be explained the sensitiveness of this measure to community sizes. *Leading Eigenvector* is clearly the less performing algorithm exhibiting a reverse sigmoid shape behavior. Even for small values of the mixing coefficient, it cannot find the exact community structure. Performances decrease more linearly for *Commfind* and *FastGreedy*. They can discover the exact community structure only when the communities are well separated. The curve representing the performance of *Radetal* is not regular. Like the linearly decreasing algorithms, its performance is good for well separated communities but the mixing coefficient increases its performance is not monotonic, and alternates between increases and decreases. When the mixing coefficient gets closer to the limit, it undergoes a sudden drop like reverse sigmoid-shaped algorithms. If we consider its average performance, we see that *Radetal* does not perform as well as the algorithms having a reverse sigmoid shape but better than the linearly decreasing ones.

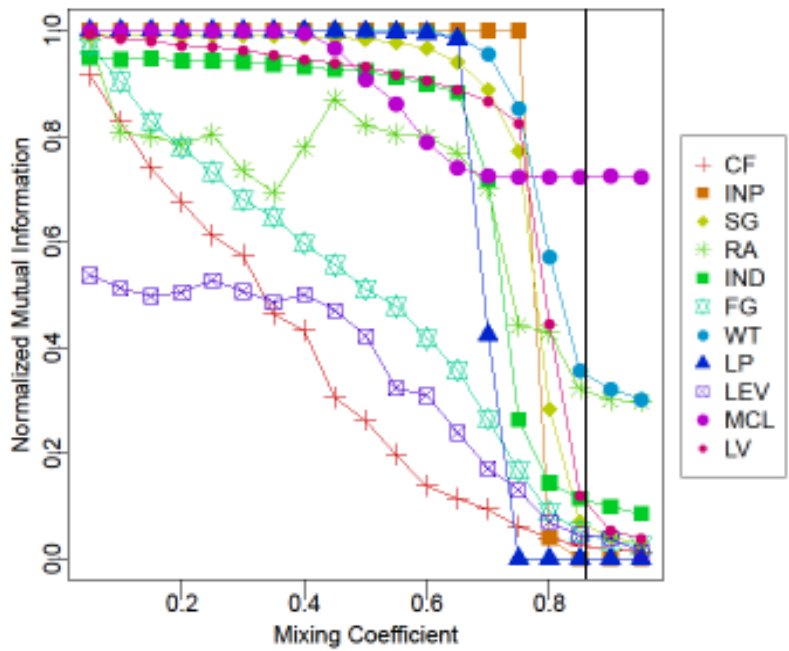

**Figure 1.** The networks are generated with parameter values $n = 5000, \gamma \approx 3, \beta = 2, \langle k \rangle = 30$. Each point corresponds to an average over 25 networks. The vertical line at $\mu = 0.86$ represents the average limit above which communities stop being clearly separated. Performances are expressed in terms of NMI for 11 algorithms.

We do not represent the standard deviation on the curves as it affects the overall readability of the figure. We nevertheless considered this information during the evaluation of the algorithms. All algorithms exhibit high standard deviation when the performance decreases. Additionally, standard deviation is overall generally higher for *Leading Eigenvector*, *Commfind*, *Radetal* and *Label Propagation*. *Label Propagation* is particularly erratic: when it is applied several times on the same network, it can output very different partitions. Besides, the density of the networks affects the consistency and performance of the algorithms too. All else being equal, the performance dispersion between sample networks increases as network density increases. We observe a similar behaviour for all algorithms when considering larger size networks. Nevertheless we observe three types of behaviour: performances can get better, worse or remains the same when the networks get larger. *Infomap* and *Walktrap* are already the best performing algorithms for smaller networks, and they perform better for larger ones. On the contrary, *Radetal, Commfind, Leading Eigenvector, Louvain* and *Infomod* perform worse. *Radetal, Commfind, Leading Eigenvector* cannot find the exact community structures and their performances start dropping before the mixing coefficient reaches the limit value ($\mu_{lim} = 0.86$). This is not as true for *Infomod* and *Louvain*, but while they both can find exact community structures for lower network sizes, they cannot do the same thing for higher ones. The other algorithms performances stay the same. Fixing all other parameters and increasing network size in LFR model increases community number and community size heterogeneity. This may explain the observed differences between the different algorithms. In fact some algorithms may be more sensitive to the size of communities.

The average degree parameter also affects the algorithms performances. Figure 2 displays the NMI values of *Walktrap* for different average degrees. We present here the most significant plot, but our comments are valid for all algorithms. For $\mu < \mu_{lim}$, we see that the higher the degree, the better the performance. This can be explained by the fact communities have a higher density due to the larger average degree. The parameter $\mu$ determines the proportion of outer links to the total degree of a node. When the average degree increases for a fixed $\mu$, the total number of inter-community and intra-community links increase. But the inter-community links are distributed between many different communities. That is why, the relative cohesion of communities increases. Thus, to discover the community structure is easier for the algorithms. However, if the network density exceeds a limit which depends on the network and community size, we cannot observe the same effect.

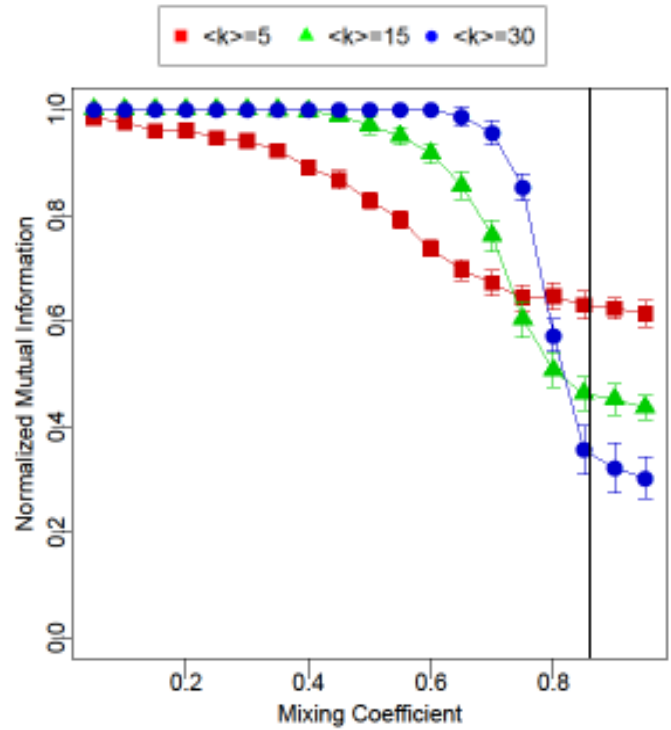

**Figure 2.** The networks are generated with parameter values $n = 5000, \gamma \approx 3, \beta = 2, \langle k \rangle = 5, 15, 30$. Each point corresponds to an average over 25 networks. The vertical line at $\mu = 0.86$ represents the average limit above which communities stop being clearly separated. Performances are expressed in terms of NMI for Walktrap algorithm.

Although it is somewhat risky to generalize from the study of a limited number of algorithms we can still learn something about the comparative performance of different categories of algorithms studied. First of all spectral and link-centrality-based algorithm's partitioning performance is one step behind than other categories. Their performance depends on the network generation parameters very much. They are not only working with high dispersion for different network sizes and average degrees but also results worse. It is difficult to form an opinion as clear-cut on the behavior of modularity optimization and information-based algorithms categories. Indeed some algorithms from those categories (*Infomod* from information-based and *FastGreedy* from modularity optimization) do not work well while others give very satisfactory results. As for random-walk based category both algorithms under study perform well for a great range of parameter variations. They exhibit less dispersion and their performance does not change very much with average degree or network size changes. *LabelPropagation* which is the only representative of the other algorithms category exhibits different performance trend depending on network size. If we do not take into account its high standard deviation we may conclude that it is more suited to large network size.

Computation time is also an important issue for practical applications. From this point of view *LabelPropagation* is the most effective. This algorithm works in linear time. Unfortunately it is characterized by a wide dispersion. Louvain and *Infomap* are not excessively time consuming. However, they still work almost %80 slower than *LabelPropagation*. Although the partitioning performances of *Walktrap*, *MarkovCluster* and *Spinglass* are good, those algorithms are very CPU intensive. This limits their use on large networks.

## 5. Conclusion

In this paper, we compared eleven different community detection algorithms. We used a set of artificial networks generated with the model defined by Lancichinetti *et al*. This model produces networks with a community structure, and power-law-distributed degree and community size. A specific parameter called mixing coefficient allows controlling the strength of the community structure. We used the normalized mutual information measure to assess the performance of the algorithms. Our results show variations of the mixing coefficient have a clear effect on the algorithms performances. Network size and average degree also affect the performances, to a lesser extent though. Other parameters of the model have a very limited effect.

In all cases *Infomap* outperforms all the other algorithms under investigation. It succeeds in identifying the communities even for high mixing coefficient values. *Walktrap*, *MarkovCluster*, *SpinGlass* and *Louvain* also have an excellent performance level, although not as good. *Leading Eigenvector* and *Commfind* are clearly outclassed. *Label Propagation* exhibits very different results for sample networks generated with the same parameters values. The network size and the average degree are two other parameters which influence algorithms performance. When the network size increases, some algorithms (*Infomap*, *Infomod*, *Louvain*) perform better, some others perform worse (*Commfind*, *SpinGlass*, *LeadingEigenvector*, *Louvain*, *Radetal*) and for the rest of the algorithms (*Walktrap*, *FastGreedy*, *MarkovCluster*), the performances does not change. For all algorithms, the higher the degree, the better the performance.

To assess the quality of community structure discovery algorithms, we rely on the normalized mutual information. However this measure, which is commonly used in the context of classical, clustering to compare two different partitions, does not take into account the topological properties of the compared community structures. A natural extension of this work consists of investigating this complementary aspect. In the future, we plan to use community-oriented topological measures to compare the community structure revealed by competing algorithms. The interest of such an approach is to shed additional light on the present evaluation.